\documentclass[aps,preprint,nofootinbib,superscriptaddress]{revtex4}%
\usepackage{amsfonts}
\usepackage{amsmath}
\usepackage{amssymb}
\usepackage[english]{babel}
\usepackage{graphicx}
\usepackage{subfig}
\usepackage{float}
\usepackage{graphicx}
\usepackage{bbm}%
\providecommand{\U}[1]{\protect\rule{.1in}{.1in}}

\begin{document}
\title{Bianchi IX cosmologies in the Einstein-Skyrme system in a sector with
non-trivial topological charge}
\author{Fabrizio Canfora}
\affiliation{Centro de Estudios Cient\'{\i}ficos (CECS), Casilla 1469,
Valdivia, Chile}
\email[]{canfora@cecs.cl}
\author{Nikolaos Dimakis}
\affiliation{Center for Theoretical Physics, College of Physical Science and Technology Sichuan University, Chengdu 610065, China}
\email[]{nsdimakis@gmail.com}
\author{Alex Giacomini}
\affiliation{Instituto de Ciencias F\'isicas y Matem\'aticas, Universidad Austral de Chile, Valdivia, Chile}
\email[]{alexgiacomini@uach.cl}
\author{Andronikos Paliathanasis}
\affiliation{Institute of Systems Science, Durban University of Technology, Durban 4000, South Africa}
\email[]{anpaliat@phys.uoa.gr}

\begin{abstract}
The dynamics of the most general Bianchi IX cosmology with three time
dependent scale factors for the Einstein-Skyrme system is analyzed. For the
Skyrmion, a generalized hedgehog ansatz with unit baryon charge is introduced.
The most remarkable feature of this ansatz is that, in the above topologically
non-trivial sector with unit topological charge, the Skyrme field equations
are identically satisfied on any Bianchi IX metric. We will show that due to
this feature the complete set of coupled Einstein-Skyrme field equations can
be deduced from a suitable minisuperspace Lagrangian. The latter allows to
perform a systematic study of the integrability properties of the
Einstein-Skyrme system for the Bianchi IX cosmology. Moreover, some analytic
and algebraic solutions for the Einstein-Skyrme model are derived. Another
remarkable consequence of the present formalism is that it is possible to
derive the Wheeler de-Witt equation for the Bianchi IX metric in the
Einstein-Skyrme cosmology in which all the effects of the Skyrmion are encoded
in an effective potential of the minisuperspace Lagrangian.

\end{abstract}
\maketitle

\section{Introduction}

It has been proposed that today's isotropic universe may have evolved from an
initial anisotropic phase. This has led to the study of a wide range of
cosmological scenarios based on a spatially homogeneous, but anisotropic
manifold. There exist nine types of space-times whose space-like surface
admits a three dimensional group of motions acting simply transitively on it
\cite{Ryan}. That is one for each three dimensional algebra spanned by the
generators of the corresponding group, the nine well known Bianchi types
\cite{Bianchi1}. Of all these models, the types I, V and IX gather the
majority of the interest in cosmology due to containing the flat, open and
closed FLRW universe respectively. Those models are used in the study of
anisotropies of the primordial universe and its evolution towards the observed
isotropy of the present epoch \cite{Mis69,jacobs2}.

Among them, the Bianchi type IX model is the most complicated and at the same
time the most interesting in many aspects. The space-time has a topology
$\mathbb{R}\times S^{3}$ with the isometry group of the three metric being the
rotation $SO(3)$ group. It contains as special cases, the FLRW metric with a
positive spatial curvature and the Taub-NUT space-time \cite{Taub,NUT}. The
study of the cosmological properties of the type IX model was initiated with
the pioneering work of Misner with his mixmaster universe \cite{Misner1} and
of course the studies of Belinskii, Khalatnikov and Lifshitz on the nature of
the approximate behavior of solutions to Einstein's equations near the initial
singularity \cite{BKL1,BKL2}.

Analytic solutions in the case where the space-time exhibits a locally
rotational symmetry (LRS) have been derived and studied both in vacuum and in
the presence of matter \cite{Taub,NUT,Ellis,Ellisbook,Steph,Barrow,mat1}. In
the vacuum case, and when the dynamics are described by two scale factors, the
system is integrable since it possess two additional independent integrals of
motion \cite{Osuga}. The situation however is not so simple when anisotropy is
introduced in all directions. The most general Bianchi type IX with three
different scale factors in the metric, apart from its complexity, has also
been the source of great controversy and debate in the literature (about both
its integrability and its chaotic behavior). The reason of controversy however
lies most of the time on the different existing notions of integrability, as
well as on the application of non-covariant criteria - that are constructed
for regular mechanical systems - over a configuration that is gauge invariant.
As far as the notion of Liouville integrability is concerned, it was found in
\cite{Cushman} and \cite{Dimnew} that the Bianchi type IX is locally
integrable, while in \cite{Llibre} it is proven that the system is not
completely integrable in terms of analytic first integrals. A vast amount of
papers (see \cite{Intp0,Intp0a,Intp1,Intp2,Intp3,Intp4,Intp5} and references
therein) is also dedicated to the study of integrability in terms of whether
the corresponding dynamical system passes the Painlev\'{e} test: the available
results in the literature are not conclusive yet.

One may wonder whether the above nice results on the Bianchi IX cosmology are
preserved if physical matter fields are included. A natural guess that Bianchi
IX metric in general relativity can be coupled self-consistently only to very
idealized matter sources. The reason is that, at a first glance, one may think
that only very simple matter fields are able to produce an energy-momentum
tensor compatible with the Bianchi IX geometries. On the other hand, strongly
interacting matter fields (such as hadronic matter and, more generically, the
low energy limit of QCD) are likely to break the properties the Bianchi IX cosmologies.

In this work we will show that this is not the case at least in the
Einstein-Skyrme model (which is the low energy limit of QCD minimally coupled
to General Relativity in the large \textbf{N} limit \cite{witten0,skyrme}).
Thus, Bianchi IX cosmologies have a much wider range of applicability than one
could think at a first glance. Moreover, we will also consider in some details
Bianchi IX cosmologies in the Einstein-non-linear sigma model. As it is well
known, Skyrme himself noticed \cite{skyrme} that the non-linear sigma model
does not possess solitonic solutions in flat, topologically trivial
(3+1)-dimensional space-times: that's why Skyrme introduced his famous Skyrme
term \cite{skyrme}! However, it is worth emphasizing that the beautiful
currents-algebraic arguments by Witten \cite{witten0} (see also
\cite{bala0,Bala1} and references therein)\ to show that the solitons of this
theory should be quantized at a semi-classical level as Fermions, and that
such a theory describe the low-energy limit of QCD do not make explicit use of
the Skyrme term itself but only of the fact that stable solitons with
non-trivial third homotopy class exist. Last but not least, when the
non-linear sigma model is minimally coupled to General Relativity, the
argument on the absence of regular topological solitons does not apply
anymore. It is thus of great theoretical interest to analyze Bianchi IX
cosmologies also in the Einstein non-linear sigma model system.

The Skyrme model is a Bosonic action for a $SU(N)$-valued scalar field (we
will consider here the $SU(2)$ case). Its solitons ---\textit{Skyrmions}---
represent Fermionic states whose topological charge is the baryon number
\cite{witten0,bala0,Bala1} (see, e.g. \cite{All,ANW,manton}). These results
have been generalized to curved space-times \cite{curved1f}. The
Einstein-Skyrme model has been very deeply analyzed as well not only due to
its relations with low energy QCD but also due to its relevance in general
relativity. For instance, black holes with a nontrivial Skyrme hair were found
using numerical tools in \cite{lucock,droz} providing the first
counterexamples to the well-known no-hair conjecture.

While the stability of the Skyrme black hole is already established \cite{droz2}
the stability of other hairy black holes is not fully understood yet \cite{Bizon:1994dh} (see also \cite{numerical1,numerical2}).
Cosmological applications of the Skyrme model have also been considered
\cite{cosmo,cosmo2}.

Due to the fact that the Skyrme field equations (especially in sectors with
non-trivial topological charge) have always been considered a very hard nut to
crack, one may think that to build analytic gravitating Skyrmions whose
geometry is of Bianchi IX type is a quite hopeless task. The reason is that,
in order to achieve such a goal, one should solve the Skyrme field equations
using an ansatz with non-vanishing topological charge on such Bianchi IX
geometries and solve, at the same time, the Einstein equations with the
corresponding Skyrme energy momentum tensor in a self-consistent way.

In fact, very recently, following the generalized hedgehog ansatz developed in
\cite{canfora2,canfora2a,canfora2b,canfora2c,yang1,ferreira,canfora9,canfora8,tubes,canfora10,Giacomini:2017xno,cawor}
the first analytic self-gravitating $SU(2)$ Skyrmions have been constructed in
\cite{canfora6,canfora6b,canfora6c}.

Here it will be shown that the gravitating topologically non-trivial
configurations found in \cite{canfora6,canfora6b,canfora6c} can be generalized
to the full Bianchi IX family with three independent scale factors. This
technique allows to construct an \textit{ansatz for the Skyrmion which
satisfies identically the Skyrme field equations on any Bianchi IX metric} in
a sector with non-vanishing topological charge.

Due to the fact that the complete set of coupled Einstein-Skyrme field
equations in the above mentioned non-trivial topological sector can be reduced
to three dynamical equations\footnote{Indeed, the generalized hedgehog ansatz
works very well also in the Bianchi IX case since the Skyrme field equations
are identically satisfied in any Bianchi IX metric as it will be explained in
the following sections.} for the three Bianchi IX scale factors, one can
derive a mini-superspace action. These results allow to discuss the classical
integrability properties of the Bianchi IX metric in the Einstein-Skyrme
system when a non-trivial topological soliton is present as source of the
Einstein equations. Another remarkable consequence of the present formalism is
that it is possible to derive the Wheeler de-Witt equation for the Bianchi IX
metric in the Einstein-Skyrme cosmology in which all the effects of the
Skyrmion are encoded in an effective potential.

The paper is organized as follows: In section \ref{Sec2} we give the general
description of the system that we study and we introduce the ansatz for both
the matter field as well as for the base manifold. In section \ref{Sec3} we
provide the equations of motion for the general system and derive the
mini-superspace Lagrangian that produces them as its Euler-Lagrange equations.
Later, in section \ref{Sec4} we examine the isotropic case and provide the
complete solution space for the FLRW universe. Section \ref{Sec5} is devoted
to the study of the LRS case with two scale factors present. In section
\ref{Sec6} we derive the Wheeler-DeWitt equations that correspond to the
previous cases and finally in section \ref{Sec7} we draw our basic conclusions
over this work.

\section{The Einstein-Skyrme System}

\label{Sec2}

We are interested in the system composed by Einstein's relativity minimally
coupled to the Skyrme action where the corresponding field is an element of
the $SU(2)$ group. Thus, our starting point is
\begin{equation}
\label{action}I =\int d^{4} x \sqrt{-g} \left[  \frac{1}{\kappa} \left(
\frac{1}{2}R + \Lambda\right)  + \frac{K}{4} \mathrm{Tr}\left(  A^{\mu}A_{\mu
}+ \frac{\lambda}{8} F_{\mu\nu}F^{\mu\nu}\right)  \right]  ,
\end{equation}
where $\kappa=8\pi G$ and $\Lambda$ are the gravitational and cosmological
constants respectively. Throughout the paper we work in units $c=\hbar=1$. The
constants $K$ and $\lambda$ are the ones corresponding to the Skyrme coupling.
The field $A_{\mu}$ is given by $A_{\mu}= U^{-1} \nabla_{\mu}U$ with $U \in
SU(2)$ and strength of the field is $F_{\mu\nu}= [A_{\mu},A_{\nu}]$. The
$SU(2)$ base in which the $A_{\mu}= A^{i}_{\mu}t_{i}$ are expanded is given by
$t_{i} = \mathbbmtt{i} \sigma_{j}$, where $\sigma_{j}$ are the Pauli matrices.
The space-time indices are denoted by Greek letters and $\mathrm{Tr}$ is the
trace over the group indices that are expressed by Latin letters.

Variation with respect to the fields $A_{\mu}$ and $g_{\mu\nu}$ in the action
integral (\ref{action}) leads the combined set of the Einstein-Skyrme
equations
\begin{subequations}
\label{eqofmo}%
\begin{align}
\nabla^{\mu}A_{\mu}+\frac{\lambda}{4}\nabla^{\mu}[A^{\nu},F_{\mu\nu}] &
=\Sigma^{j}t_{j}=0\ ,\label{Skeq}\\
G_{\mu\nu}+\Lambda g_{\mu\nu} &  =\kappa T_{\mu\nu}\ ,\label{Eineq}%
\end{align}
where $\Sigma^{j}=0$ ($j$= 1, 2, 3) are the Skyrme field equations, $G_{\mu
\nu}=R_{\mu\nu}-\frac{1}{2}g_{\mu\nu}R$ is the Einstein tensor and
\end{subequations}
\begin{equation}
T_{\mu\nu}=-\frac{K}{2}\mathrm{Tr}\left[  A_{\mu}A^{\mu}-\frac{1}{2}g_{\mu\nu
}A^{\kappa}A_{\kappa}+\frac{\lambda}{4}\left(  g^{\kappa\lambda}F_{\mu\kappa
}F_{\nu\lambda}-\frac{1}{4}g_{\mu\nu}F_{\kappa\lambda}F^{\kappa\lambda
}\right)  \right]  \label{enmom}%
\end{equation}
is the energy momentum tensor of the Skyrmion.

\subsection{The generalized Hedgehog ansatz}

We adopt the standard parametrization for the $SU(2)$ scalar $U(x^{\mu})$:
\begin{equation}
U^{\pm1}=Y^{0}(x^{\mu})\mathbb{I}\pm Y^{i}(x^{\mu})t_{i},\quad(Y^{0}%
)^{2}+(Y^{i})^{2}=1 \label{bans0}%
\end{equation}
where $\mathbb{I}$ is the two dimensional identity matrix and $Y^{\mu}$ is
parametrized as
\begin{equation}
Y^{0}=\cos\alpha,\quad Y^{i}=n^{i}\sin\alpha\label{bans1}%
\end{equation}
and
\begin{equation}
n^{1}=\sin\Theta\cos\Phi,\quad n^{2}=\sin\Theta\sin\Phi,\quad n^{3}=\cos
\Theta. \label{bans2}%
\end{equation}

In what follows we denote the space-time coordinates with the variables
$x^{\mu}=(t,\theta,\phi,\gamma)$ and we adopt the following ansatz for the
Skyrmion (see \cite{canfora6,canfora6b,canfora6c} and references therein)
\begin{equation}
\Phi=\frac{\gamma+\phi}{2},\quad\tan\Theta=\frac{\cot\left(  \frac{\theta}%
{2}\right)  }{\cos\left(  \frac{\gamma-\phi}{2}\right)  },\quad\tan
\alpha=\frac{\sqrt{1+\tan^{2}\Theta}}{\tan\left(  \frac{\gamma-\phi}%
{2}\right)  }\ , \label{bSkans}%
\end{equation}%
\begin{equation}
R_{\mu}=U^{-1}\partial_{\mu}U=R_{\mu}^{i}t_{i}\ .\label{bSkans2}%
\end{equation}
The explicit expression for the $R_{\mu}$ can be found in appendix
\ref{appRmu}.

In Eq. (\ref{bSkans}), the range of the coordinates $\theta$, $\gamma$ and
$\phi$ is
\begin{equation}
\label{rangevar}0 \leq\theta\leq\pi, \quad0 \leq\gamma\leq4\pi, \quad0
\leq\phi\leq2\pi.
\end{equation}

The remarkable properties of the above ansatz will be fully apparent after we
will introduce the Bianchi IX metric and we will discuss the corresponding
Skyrme field equations in the next sections.

In the Skyrme theory, the Baryon charge reads
\begin{equation}
W=B=\frac{1}{24\pi^{2}}\int_{\left\{  t=const\right\}  }\rho_{B}\ ,
\label{rational4}%
\end{equation}%
\begin{equation}
\rho_{B}=\epsilon^{ijk}Tr\left(  U^{-1}\partial_{i}U\right)  \left(
U^{-1}\partial_{j}U\right)  \left(  U^{-1}\partial_{k}U\right)  \ .
\label{rational4.1}%
\end{equation}

In terms of $\alpha$, $\Theta$ and $\Phi$,\ the topological density $\rho_{B}$
is written as
\begin{equation}
\rho_{B}=12\left(  \sin^{2}\alpha\sin\Theta\right)  d\alpha\wedge
d\Theta\wedge d\Phi= \frac{3}{2} \sin\theta\, d\theta\wedge d\phi\wedge
d\gamma, \label{rational4.1.1}%
\end{equation}
so that a necessary condition in order to have non-trivial topological charge
is%
\begin{equation}
d\alpha\wedge d\Theta\wedge d\Phi\neq0\ . \label{necscond}%
\end{equation}

From the geometrical point of view the above condition (which simply states
that $\alpha$, $\Theta$ and $\Phi$ must be three independent functions) can be
interpreted as saying that such three functions ``fill a three-dimensional
spatial volume" at least locally. In other words, $d\alpha$, $d\Theta$ and
$d\Phi$ can be used as 3D volume form. Hence, the condition in Eq.
(\ref{necscond}) ensures that the configuration one is interested in describes
a genuine three-dimensional object. In the case of the ansatz defined in Eqs.
(\ref{bans1}), (\ref{bans2}) and (\ref{bSkans}) a direct computation shows
that
\begin{equation}
\frac{1}{24\pi^{2}}\int_{\left\{  t=const\right\}  }\rho_{B}=1.
\label{necscond2}%
\end{equation}
Usually, this second requirement allows to fix some of the parameters of the ansatz.

From now on, as it is customary in the literature, the terms
\textit{Gravitating Skyrmion} will refer to smooth regular solutions of Eqs.
(\ref{eqofmo}) with the properties that the topological charge (defined in
Eqs. (\ref{rational4}), (\ref{rational4.1}), (\ref{rational4.1.1}) and
(\ref{necscond})) is non-vanishing.

\subsection{Metric ansatz}

The general spatially homogeneous Bianchi type IX space-time admits a three
dimensional Killing algebra with a fully anisotropic scale factor matrix
\begin{equation}
h_{\alpha\beta}=%
\begin{pmatrix}
a(t)^{2} & 0 & 0\\
0 & b(t)^{2} & 0\\
0 & 0 & c(t)^{2}%
\end{pmatrix}
. \label{scfmat}%
\end{equation}

It is known that the line element for the underlying geometry can assume the
general form
\begin{equation}
ds^{2}=-N(t)^{2}dt^{2}+h_{\alpha\beta}(t)\omega_{i}^{\alpha}(x)\omega
_{i}^{\beta}(x)dx^{i}dx^{j},\quad i,j,\alpha,\beta=1,2,3, \label{lineel}%
\end{equation}
where $N(t)$ is the lapse function and the $\omega_{\alpha}$'s are the 1-forms
corresponding to the invariant basis of the three dimensional surface,
characterized by the three dimensional group of isometries\footnote{In this
section we must be careful so that there is no confusion with respect to the
indices: Greek letters do not denote space-time indices, but they refer to the
coefficients of the scale factor matrix $h_{\alpha\beta}$. Moreover, the Latin
indices $i,j$ count the three spatial dimensions and do not denote the $SU(2)$
group indices that we saw previously.}.

A more general line element, involving possibly a non zero shift vector field,
can always be brought to this form (at least locally) by invoking
time-dependent automorphisms of the algebra of the invariant basis that
correspond to space-time diffeomorphisms \cite{tchris}.

In our case the spatial coordinate are denoted by $x=(\theta,\phi,\gamma)$ and
for the one forms $\omega$ we consider \cite{Ryan}
\begin{subequations}
\label{Cartanforms}%
\begin{align}
\omega^{1}  &  =-\sin\gamma\,d\theta+\sin\theta\cos\gamma\,d\phi
\\
\omega^{2}  &  =\cos\gamma\,d\theta+\sin\theta\sin\gamma\,d\phi\\
\omega^{3}  &  =\cos\theta\,d\phi+d\gamma.
\end{align}

Moreover, one can directly see that the $\omega^{j}$ defined above satisfy the
relation
\end{subequations}
\begin{equation}
d\omega^{\alpha}=\frac{1}{2}C_{\beta\gamma}^{\alpha}\omega^{\beta}\wedge
\omega^{\gamma},
\end{equation}
with the structure constants given in terms of the Levi-Civita symbol in three
dimensions $C_{\beta\gamma}^{\alpha}=\epsilon_{\alpha\beta\gamma}$ (we assume
$\epsilon_{123}=+1$).

We observe that in the general case the three dimensional part of the metric
depends - apart from time - on two spatial variables $\gamma$ and $\theta~$as
follows
\begin{equation}%
\begin{split}
ds^{2}=  &  -N^{2}dt^{2}+\left(  a^{2}\sin^{2}\gamma+b^{2}\cos^{2}%
\gamma\right)  d\theta^{2}+c^{2}d\gamma^{2}\\
&  +\left[  \left(  a^{2}\cos^{2}\gamma+b^{2}\sin^{2}\gamma\right)  \sin
^{2}\theta+c^{2}\cos^{2}\theta\right]  d\phi^{2}\\
&  +2c^{2}\cos\theta d\gamma d\phi+\left(  b^{2}-a^{2}\right)  \sin
(2\gamma)\sin\theta d\theta d\phi
\end{split}
. \label{lineelexpl}%
\end{equation}

On the other hand, when one considers the LRS case, where only two of the
scale factors are independent, i.e. $a(t)=b(t)$, then only one spatial
variable remains in the final form for the line element since
\eqref{lineelexpl} reduces to
\begin{equation}
ds^{2}=-N^{2}dt^{2}+b^{2}d\theta^{2}+c^{2}d\gamma^{2}+\left(  b^{2}\sin
^{2}\theta+c^{2}\cos^{2}\theta\right)  d\phi^{2}+2c^{2}\cos\theta d\gamma
d\phi. \label{lineelLRS}%
\end{equation}

An analysis involving solutions of the Einstein-Skyrme model under this latter
ansatz for the line element has been given in \cite{FBZ,FPTZ}.

The main property of the Skyrme ansatz in Eqs. (\ref{bans0}) - (\ref{bSkans2})
is that the $SU(2)$ left-invariant one-forms constructed from it essentially
coincide with the spatial drei-beins used to build the Bianchi IX metric in
Eq. \eqref{Cartanforms} (see the appendix \ref{appRmu}). In particular, due to this
fact, whenever a spatial drei-bein is contracted with the left-invariant forms
arising from the Skyrmion Kronecker delta terms arise and this leads to
considerable simplifications in the Skyrme field equations.

In what follows, we adopt the Misner variables $(\Omega,\beta_{+},\beta_{-})$
that are associated to the scale factors $(a,b,c)$ through the change of
variables
\begin{equation}
a=e^{\beta_{+}+\sqrt{3}\beta_{-}-\Omega},\quad b=e^{\beta_{+}-\sqrt{3}%
\beta_{-}-\Omega},\quad c=e^{-2\beta_{+}-\Omega}.\label{Misnvar}%
\end{equation}
In this parametrization of the configuration space variables, the ensuing
mini-superspace metric assumes a simple diagonal form.

\section{Equations of motion and mini-superspace Lagrangian}

\label{Sec3}

\subsection{Einstein's Equations}

By adopting the previously discussed choices for the Skyrme field
\eqref{bSkans} and for a space-time metric \eqref{lineelexpl} into the field
equations \eqref{eqofmo} it can be straightforwardly verified that the three
equations \eqref{Skeq} (one for each $t_{i}$) are satisfied identically. At
the same time, \eqref{Eineq} reduce to a set of ordinary differential
equations, which we denote with
\begin{equation}
E_{\mu\nu}:=G_{\mu\nu}+\Lambda g_{\mu\nu}-\kappa T_{\mu\nu}=0.
\end{equation}

By using \eqref{Misnvar}, we can write the following system:
\begin{equation}%
\begin{split}
&  \frac{\bar{K}}{8}\Big(e^{-2\beta_{+}+2\sqrt{3}\beta_{-}+2\Omega}%
+e^{-2\beta_{+}-2\sqrt{3}\beta_{-}+2\Omega}-2e^{4\beta_{+}+2\Omega}\Big)\\
&  -\frac{\bar{\lambda}}{32}\left(  e^{2\beta_{+}-2\sqrt{3}\beta_{-}+4\Omega
}+e^{2\beta_{+}+2\sqrt{3}\beta_{-}+4\Omega}-2e^{4\Omega-4\beta_{+}}\right)  \\
&  +e^{2\Omega-8\beta_{+}}-\frac{1}{2}\left(  e^{4\beta_{+}-4\sqrt{3}\beta
_{-}+2\Omega}+e^{4\beta_{+}+4\sqrt{3}\beta_{-}+2\Omega}\right)  +\frac
{3\dot{N}\dot{\beta}_{+}}{N^{3}}+\frac{9\dot{\beta}_{+}\dot{\Omega}}{N^{2}%
}-\frac{3\ddot{\beta}_{+}}{N^{2}}=0,
\end{split}
\label{spaeq1}%
\end{equation}%
\begin{equation}%
\begin{split}
&  \frac{\bar{K}}{4}\left(  e^{-2\beta_{+}+2\sqrt{3}\beta_{-}+2\Omega
}-e^{-2\beta_{+}-2\sqrt{3}\beta_{-}+2\Omega}\right)  +\frac{\bar{\lambda}}%
{16}\left(  e^{2\beta_{+}+2\sqrt{3}\beta_{-}+4\Omega}-e^{2\beta_{+}-2\sqrt
{3}\beta_{-}+4\Omega}\right)  \\
&  e^{4\beta_{+}+4\sqrt{3}\beta_{-}+2\Omega}-e^{4\beta_{+}-4\sqrt{3}\beta
_{-}+2\Omega}-\frac{2\sqrt{3}\dot{N}\dot{\beta}_{-}}{N^{3}}-\frac{6\sqrt
{3}\dot{\Omega}\dot{\beta}_{-}}{N^{2}}+\frac{2\sqrt{3}\ddot{\beta}_{-}}{N^{2}%
}=0,
\end{split}
\label{spaeq2}%
\end{equation}%
\begin{equation}%
\begin{split}
&  3\Lambda+\frac{\bar{K}}{8}\left(  e^{-2\beta_{+}-2\sqrt{3}\beta_{-}%
+2\Omega}+e^{-2\beta_{+}+2\sqrt{3}\beta_{-}+2\Omega}+e^{4\beta_{+}+2\Omega
}\right)  \\
&  -\frac{\bar{\lambda}}{32}\left(  e^{2\left(  \beta_{+}+\sqrt{3}\beta
_{-}+2\Omega\right)  }+e^{2\beta_{+}-2\sqrt{3}\beta_{-}+4\Omega}%
+e^{4\Omega-4\beta_{+}}\right)  \\
&  +\frac{1}{4}\left(  e^{4\beta_{+}-4\sqrt{3}\beta_{-}+2\Omega}+e^{4\beta
_{+}+4\sqrt{3}\beta_{-}+2\Omega}+e^{2\Omega-8\beta_{+}}\right)  \\
&  -9\left(  \frac{\dot{\beta}_{+}^{2}}{N^{2}}+\frac{\dot{\beta}_{+}^{2}%
}{N^{2}}+\frac{\dot{\Omega}^{2}}{N^{2}}\right)  -\frac{6\dot{N}\dot{\Omega}%
}{N^{3}}+\frac{6\ddot{\Omega}}{N^{2}}=0
\end{split}
\label{spaeq3}%
\end{equation}
and
\begin{equation}%
\begin{split}
&  \Lambda+\frac{\bar{K}}{8}\left(  e^{-2\beta_{+}-2\sqrt{3}\beta_{-}+2\Omega
}+e^{-2\beta_{+}+2\sqrt{3}\beta_{-}+2\Omega}+e^{2\Omega+4\beta_{+}}\right)  \\
&  +\frac{\bar{\lambda}}{32}\left(  e^{2\left(  \beta_{+}+\sqrt{3}\beta
_{-}+2\Omega\right)  }+e^{2\beta_{+}-2\sqrt{3}\beta_{-}+4\Omega}%
+e^{4\Omega-4\beta_{+}}\right)  \\
&  +\frac{1}{4}\left(  e^{4\beta_{+}-4\sqrt{3}\beta_{-}+2\Omega}+e^{4\beta
_{+}+4\sqrt{3}\beta_{-}+2\Omega}+e^{2\Omega-8\beta_{+}}\right)  +\frac
{3}{N^{2}}\left(  \dot{\beta}_{+}^{2}+\dot{\beta}_{-}^{2}-\dot{\Omega}%
^{2}\right)  =0,
\end{split}
\label{con}%
\end{equation}
where $\bar{K}=\kappa K-4$ and $\bar{\lambda}=\kappa K\lambda$.

Thus, as promised, the complete set of coupled Einstein-Skyrme field equations
in a sector with non-vanishing Baryon charge reduce to three dynamical
equations for the three Bianchi IX scale factors and a constraint. This is a
quite important technical achievement which opens the possibility to
generalize many of the known results in the literature on Bianchi IX cosmology
in situations in which a topological soliton is consistently coupled to
General Relativity.

When $\kappa K=4$ (so that $\bar{K}=0$ in the above equations) and $\lambda$
vanishes (which corresponds to Einstein non-linear sigma model system for a
particular value of the coupling constant) the corresponding field equations
have very special properties: this issue will be discussed in the following sections.

Each of \eqref{spaeq1}-\eqref{con} corresponds to a combination of Einstein's
equations so that the second derivatives are isolated in each one of them. The
relations that lead to \eqref{spaeq1}-\eqref{con} are respectively
\begin{align}
&  E_{\;3}^{3}-E_{\;2}^{2}-\frac{1}{2}\cot(2\gamma)\frac{\partial E_{\;2}^{2}%
}{\partial\gamma}=0\\
&  \frac{1}{\sin(2\gamma)}\frac{\partial E_{\;1}^{1}}{\partial\gamma}=0\\
&  E_{\;3}^{3}+2E_{\;2}^{2}+\cot(2\gamma)\frac{\partial E_{\;2}^{2}}%
{\partial\gamma}=0\\
&  E_{\;0}^{0}=0.
\end{align}

The fact that no other equation regarding the matter degrees of freedom
appears, it is owed to the clever selection of an ansatz for the Skyrme field,
\eqref{bSkans}, that makes the relevant equations satisfied identically.
Therefore, we are left with a system consisting of three second-order,
ordinary differential equations (ODEs) and a constraint equation involving
only first-order derivatives. This means that we have just two physical
degrees of freedom; for a formal counting of the physical degrees of freedom
in constrained systems we refer the reader to \cite{Henneaux,Diaz}. Hence, it
is expected that the system can be reduced - upon satisfaction of the
constraint - to two second order (non autonomous in general) ODEs.

One way to do so is by solving the constraint equation \eqref{con}
algebraically with respect to the lapse $N$ and substitute to the rest three
equations \eqref{spaeq1}-\eqref{spaeq3}. At that point, the resulting system
can be solved algebraically with respect to just two of the three
accelerations involved. The third corresponds to a gauge degree of freedom.
The ensuing equations are extremely complicated. However, in the special case
where $\bar{\lambda}=\Lambda=0$ it is interesting to note that a Lie-point
symmetry is present, which also exists for the vacuum case. It is worth to
remark here that such a case is still very interesting since it corresponds to
the Einstein non-linear sigma model system. The symmetry is broken by the
general Skyrme field, but under the previous choice for the parameters it is reinstated.

So, for the particular case $\bar{\lambda}=\Lambda=0$, if we follow this
recipe: a) solve the constraint \eqref{con} algebraically for $N$, b)
substitute into \eqref{spaeq1}-\eqref{spaeq3} and c) choose the redundant
degree of freedom to be some function of time $t$ i.e. fix the gauge (in our
case we use $\Omega(t)=t$), then we obtain the system
\begin{equation}%
\begin{split}
\ddot{\beta}_{+}=  &  \bigg[\frac{-3\bar{K}\left(  e^{4\sqrt{3}\beta_{-}%
}+1\right)  e^{6\beta_{+}+2\sqrt{3}\beta_{-}}-12e^{4\sqrt{3}\beta_{-}}}%
{\bar{K}e^{6\beta_{+}+6\sqrt{3}\beta_{-}}+\bar{K}e^{6\beta_{+}+2\sqrt{3}%
\beta_{-}}+e^{4\sqrt{3}\beta_{-}}\left(  \bar{K}e^{12\beta_{+}}+2\right)
+2e^{12\beta_{+}+8\sqrt{3}\beta_{-}}+2e^{12\beta_{+}}}\\
&  -2\dot{\beta}_{+}+2\bigg]\left(  \dot{\beta}_{+}^{2}+\dot{\beta}_{-}%
^{2}-1\right)
\end{split}
\label{eqb1}%
\end{equation}%
\begin{equation}%
\begin{split}
\ddot{\beta}_{-}=  &  \Bigg[\frac{\sqrt{3}e^{6\beta_{+}}\left(  e^{4\sqrt
{3}\beta_{-}}-1\right)  \left(  \bar{K}e^{2\sqrt{3}\beta_{-}}+4e^{6\beta
_{+}+4\sqrt{3}\beta_{-}}+4e^{6\beta_{+}}\right)  }{\bar{K}e^{6\left(
\beta_{+}+\sqrt{3}\beta_{-}\right)  }+\bar{K}e^{6\beta_{+}+2\sqrt{3}\beta_{-}%
}+e^{4\sqrt{3}\beta_{-}}\left(  \bar{K}e^{12\beta_{+}}+2\right)
+2e^{12\beta_{+}+8\sqrt{3}\beta_{-}}+2e^{12\beta_{+}}}\\
&  -2\dot{\beta}_{-}\Bigg]\left(  \dot{\beta}_{+}^{2}+\dot{\beta}_{-}%
^{2}-1\right)
\end{split}
\label{eqb2}%
\end{equation}
which has the obvious Lie point symmetry generator $\partial_{t}$ since it is
autonomous. We have to note that here $1-\dot{\beta}_{+}^{2}-\dot{\beta}%
_{-}^{2}\neq0$ or else the lapse function becomes zero.

\subsection{The mini-superspace Lagrangian}

As far as the general system is concerned, if one substitutes the ansatz for
the line element and the Skyrme field in the original action \eqref{action}
and keep only the dynamical degrees of freedom we can obtain the
mini-superspace Lagrangian
\begin{equation}
L=\frac{3e^{-3\Omega}}{\kappa N}\left(  \dot{\beta}_{+}^{2}+\dot{\beta}%
_{-}^{2}-\dot{\Omega}^{2}\right)  -NV(\beta_{+},\beta_{-},\Omega
)\label{miniLag}%
\end{equation}
where
\begin{equation}%
\begin{split}
V(\beta_{+},\beta_{-},\Omega)= &  \frac{1}{\kappa}\Bigg[\frac{1}{4}\left(
e^{4\beta_{+}-4\sqrt{3}\beta_{-}-\Omega}+e^{4\beta_{+}+4\sqrt{3}\beta
_{-}-\Omega}+e^{-8\beta_{+}-\Omega}\right)  \\
&  +\frac{\bar{K}}{8}\left(  e^{-2\beta_{+}-2\sqrt{3}\beta_{-}-\Omega
}+e^{-2\beta_{+}+2\sqrt{3}\beta_{-}-\Omega}+e^{4\beta_{+}-\Omega}\right)  \\
&  +\frac{\bar{\lambda}}{32}\left(  e^{2\beta_{+}+2\sqrt{3}\beta_{-}+\Omega
}+e^{2\beta_{+}-2\sqrt{3}\beta_{-}+\Omega}+e^{\Omega-4\beta_{+}}\right)
+\Lambda e^{-3\Omega}\Bigg]
\end{split}
\end{equation}
is the potential function of the system.

It can be easily verified that this Lagrangian reproduces correctly the
dynamical evolution of the corresponding gravitational system, i.e. its
Euler-Lagrange equations are equivalent to \eqref{spaeq1} - \eqref{con}. The
$\bar{K}=-4$, $\bar{\lambda}=0$ case obviously corresponds to the pure Bianchi
IX model with a cosmological constant. Cosmological Lagrangians of this form
can be associated with pseudo-Euclidean generalized Toda systems
\cite{Ivash1,Ivash2,Gavr1}. Euclidean Toda systems are extensively studied in
the literature, but not so many results are available in the pseudo-Euclidean
case. However, there exist certain conditions under which a system of this
form can be characterized as integrable \cite{Gavr1}. This is not the case for
the fully anisotropic Bianchi type IX in vacuum and as well for the more
general system described by \eqref{miniLag}. Nevertheless, for the vacuum
case, it has been shown in \cite{Dimnew} that one can put in use non-local
conserved charges to show that enough independent commuting phase space
functions exist so as to characterize the system (at least locally) as
Liouville integrable. Something which is in accordance with the result of
\cite{Cushman}.

Unfortunately, the explicit form of these functions cannot be known (for all
of them) without the solution of the original system. Thus, making their
existence of little practical use. The same logic can be followed here to see
that the matter content that we assume does not affect this property. The
existence of the non-local conserved charges (and their Poisson algebra) is
associated with conformal Killing vectors (and their Lie algebra) of the
mini-superspace metric that we can read out of the Lagrangian \eqref{miniLag}.
The latter - due to the adopted ansatz for the Skyrme field - has the same
kinetic term as the vacuum case. Hence, the existence of the same number of
non-local conserved charges is guaranteed, even though their form will be
different due to the change of the potential part. Later in our analysis, we
use such a non-local charge in a special case of the system, in order to
perform a convenient reduction for our study.

\section{The isotropic case $\beta_{+}=\beta_{-}=0$}

\label{Sec4}

In this case, where the scale factor matrix \eqref{scfmat} is isotropic and
the corresponding mini-superspace Lagrangian reads
\begin{equation}
L=-\frac{3e^{-3\Omega}\dot{\Omega}^{2}}{\kappa N}-\frac{N}{\kappa}\left(
\frac{3}{8}(\bar{K}+2)e^{-\Omega}+\frac{3\bar{\lambda}e^{\Omega}}{32}+\Lambda
e^{-3\Omega}\right)  .\label{Lag1sc}%
\end{equation}
The corresponding field equations just reduce to
\begin{align} \label{con1sc}
&  \frac{3}{8}(\bar{K}+2)e^{2\Omega}+\Lambda+\frac{3}{32}\bar{\lambda
}e^{4\Omega}-\frac{3\dot{\Omega}^{2}}{N^{2}}=0 \\ \label{feq1sc}
&  N^{3}\left(  4(\bar{K}+2)e^{2\Omega}+32\Lambda-\bar{\lambda}e^{4\Omega
}\right)  -64\dot{N}\dot{\Omega}+N\left(  64\ddot{\Omega}-96\dot{\Omega}%
^{2}\right)  =0
\end{align}

This is a system of zero physical degrees of freedom or a pure gauge system.
We have one scale factor and one constraint equation, thus the dimension of
the reduced physical space is $1-1=0$. This means that the solution can be
obtained without integration, but simply algebraically, by solving
\eqref{con1sc} with respect to $N$, which yields
\begin{equation}
N=\frac{4\dot{\Omega}}{\left(  2(\bar{K}+2)e^{2\Omega}+\frac{1}{2}\bar
{\lambda}e^{4\Omega}+\frac{16\Lambda}{3}\right)  ^{1/2}}. \label{isoN}%
\end{equation}
Substitution of \eqref{isoN} into the remaining spatial equation for $\Omega$,
\eqref{feq1sc}, leads to the latter being satisfied identically. The line
element that corresponds to this solution is:
\begin{equation}
ds^{2}=-\frac{16}{2(\bar{K}+2)e^{2\Omega}+\frac{1}{2}\bar{\lambda}e^{4\Omega
}+\frac{16\Lambda}{3}}d\Omega^{2}+e^{-2\Omega}\left(  d\gamma^{2}+d\theta
^{2}+d\phi^{2}+2\cos\theta d\gamma d\phi\right)  , \label{soliso1}%
\end{equation}
where we see that the function $\Omega$, which remains arbitrary in the
solution, is effectively converted into the time variable.

Solution \eqref{soliso1} can of course be expressed in a cosmological time,
$\tau$, $N(\tau)=1$ gauge if we perform the transformation $\Omega\mapsto\tau$
for which
\begin{equation}
\int N(t)dt=\tau\Rightarrow\int\frac{4}{\left(  2(\bar{K}+2)e^{2\Omega}%
+\frac{1}{2}\bar{\lambda}e^{4\Omega}+\frac{16\Lambda}{3}\right)  ^{1/2}%
}d\Omega=\tau. \label{trton1}%
\end{equation}
Under transformation \eqref{trton1} - and when $\Lambda\neq0$ - the resulting
line element reads
\begin{equation}
ds^{2}=-d\tau^{2}+\frac{1}{32\Lambda}\left[  A_{0}\cosh\left(  \frac
{2\sqrt{\Lambda}\tau}{\sqrt{3}}\right)  -6(\Bar{K}+2)\right]  \left(
d\gamma^{2}+d\theta^{2}+d\phi^{2}+2\cos\theta d\gamma d\phi\right)  ,
\label{metis2}%
\end{equation}
where $A_{0}=\pm2\left(  9(\bar{K}+2)^{2}-24\bar{\lambda}\Lambda\right)
^{1/2}$. The cosmological constant in \eqref{metis2} can be either positive or
negative, choosing of course appropriately the domain of definition of the
variable $\tau$ and of the parameters, so that the signature of the metric
remains $(-,+,+,+)$.

In the $\Lambda=0$, $\bar{K}\neq-2$ case, transformation \eqref{trton1} leads
to the simple expression
\begin{equation}
ds^{2}=-d\tau^{2}+\left(  \frac{1}{8}(\bar{K}+2)\tau^{2}-\frac{\bar{\lambda}%
}{4(\bar{K}+2)}\right)  \left(  d\gamma^{2}+d\theta^{2}+d\phi^{2}+2\cos\theta
d\gamma d\phi\right)  . \label{metis3}%
\end{equation}

Finally, if $\Lambda=0$ and $\bar{K}=-2$ we find the line-element
\begin{equation}
ds^{2}=-d\tau^{2}+\frac{\bar{\lambda}^{1/2}}{2\sqrt{2}}\tau\left(  d\gamma
^{2}+d\theta^{2}+d\phi^{2}+2\cos\theta d\gamma d\phi\right)  , \label{metis4}%
\end{equation}
with the scale factor being linear in $\tau$. Solutions \eqref{metis2} and
\eqref{metis3} where both originally found and analyzed in \cite{FPTZ}.

For line element \eqref{metis2} we calculate the Ricci scalar, which reads
\begin{equation}
R=\frac{4\Lambda\left(  A_{0}\cosh\left(  \frac{2\sqrt{\Lambda}\tau}{\sqrt{3}%
}\right)  +12\right)  }{A_{0}\cosh\left(  \frac{2\sqrt{\Lambda}\tau}{\sqrt{3}%
}\right)  -6(\bar{K}+2)}.
\end{equation}

From the latter we observe that a curvature singularity is avoided when
$\Lambda>0$, if $\bar{K}<-2$ and $\Lambda\leq\frac{3(\bar{K}+2)^{2}}%
{8\bar{\lambda}}$. Also requiring of course that the space-time metric has a
Lorentzian signature. On the contrary, when $\Lambda<0$ and the dependence on
$\tau$ is periodic, there is no appropriate range of values for the parameters
so that a singularity can be avoided (given of course that due to the Skyrme
coupling we need to have $\bar{\lambda}>0$). However when $\Lambda<0$, then we
have a cyclic solution of the form $a\left(  \tau\right)  =a_{0}+a_{1}\cos\left(
\frac{2\sqrt{|\Lambda|}\tau}{\sqrt{3}}\right)  $, around the Einstein static
solution $a_{0}$. Solutions of that form have been derived and studied in the
case of closed Friedmann--Lema\^{\i}tre--Robertson--Walker metric and in the
presence of ghost fields, for more details see \cite{bar1,bar2}.

The scalar curvatures for the space-times characterized by \eqref{metis3} and
\eqref{metis4} are
\begin{equation}
R_{\Lambda=0} = \frac{6 (\bar{K}+2) (\bar{K}+4)}{(\bar{K}+2)^{2} \tau^{2}-2
\bar{\lambda} }%
\end{equation}
and
\begin{equation}
R_{\Lambda=0,\bar{K}=-2} = \frac{3 \sqrt{2}}{\bar{\lambda}^{1/2} \tau},
\end{equation}
respectively. From the above expressions we can observe the existence of a
curvature singularity at a finite time in the first case and at the origin
$\tau=0$ at the second.

\section{The LRS case}

\label{Sec5}

Here we study configurations where at least some isometry is present in the model.

\subsection{A static solution when $\beta_{+} = \pm\frac{1}{\sqrt{3}}\beta
_{-}$}

In the general case where we have all the parameters present, we can derive a
simple static solution by assuming that $\beta_{+}$, $\beta_{-}$ and $\Omega$
are all constants. Under this condition it is easy to derive
\begin{subequations}
\label{stat1}%
\begin{align}
&  \Lambda= \frac{1}{4} \left(  e^{4 \sqrt{3} \beta_{-}}+2\right)  ^{2}
e^{\frac{4 \beta_{-}}{\sqrt{3}}+2 \Omega}\\
&  \lambda= 8 \left(  e^{4 \sqrt{3} \beta_{-}}+2\right)  e^{\frac{8 \beta_{-}%
}{\sqrt{3}}-2 \Omega}\\
&  \bar{K} = -2 \left(  4 e^{4 \sqrt{3} \beta_{-}}+e^{8 \sqrt{3} \beta_{-}%
}+2\right) \\
&  \beta_{+} = \frac{1}{\sqrt{3}}\beta_{-}%
\end{align}
and
\end{subequations}
\begin{subequations}
\begin{align}
&  \Lambda= \frac{1}{4} \left(  2 e^{4 \sqrt{3} \beta_{-}}+1\right)  ^{2} e^{2
\Omega-\frac{28 \beta_{-}}{\sqrt{3}}}\\
&  \lambda= 8 \left(  2 e^{4 \sqrt{3} \beta_{-}}+1\right)  e^{-\frac{20
\beta_{-}}{\sqrt{3}}-2 \Omega}\\
&  \bar{K} = -2 e^{-8 \sqrt{3} \beta_{-}} \left(  4 e^{4 \sqrt{3} \beta_{-}}+2
e^{8 \sqrt{3} \beta_{-}}+1\right) \\
&  \beta_{+} = -\frac{1}{\sqrt{3}}\beta_{-},
\end{align}
that solve the field equations \eqref{spaeq1}-\eqref{con}. They are both cases
where there exists a local rotational symmetry. The first set corresponds to
$b=c$ in the scale factor matrix \eqref{scfmat}, while the second to $a=c$.
The constants $\beta_{-}$ and $\Omega$, with respect to which all of the rest
are parameterized, remain arbitrary in the solution. What is more, it can be
seen that they are both essential for the geometry i.e. they cannot be
absorbed by a space-time diffeomorphism. The latter becomes evident by
studying the curvature scalars. For example, for the first set \eqref{stat1}
it is easy to derive that the Ricci and Kretschmann $\mathcal{K}%
=R_{\kappa\lambda\mu\nu}R^{\kappa\lambda\mu\nu}$ scalars are
\end{subequations}
\begin{equation}
R = -\frac{1}{2} \left(  e^{4 \sqrt{3} \beta_{-}}-4\right)  e^{\frac{4
\beta_{-}}{\sqrt{3}}+2 \Omega} \quad\text{and} \quad\mathcal{K} = \frac{1}{4}
\left(  -24 e^{4 \sqrt{3} \beta_{-}}+11 e^{8 \sqrt{3} \beta_{-}}+16\right)
e^{\frac{8 \beta_{-}}{\sqrt{3}}+4 \Omega}%
\end{equation}
respectively. Obviously, the last two relations are solvable with respect to
$\beta_{-}$ and $\Omega$, hence the two constants are essential for the
geometry and cannot be absorbed inside the metric. The corresponding line
elements are given by \eqref{lineelexpl} with $a=e^{\frac{4 \beta_{-} }%
{\sqrt{3}}- \Omega}$, $b=c=e^{-\frac{2 \beta_{-}}{\sqrt{3}}-\Omega}$ for the
first case and $b= e^{-\frac{4 \beta_{-} }{\sqrt{3}}- \Omega}$, $a=c=e^{\frac
{2 \beta_{-}}{\sqrt{3}}-\Omega}$ for the second.

\subsection{The equivalent mini-superspace system for $\beta_{-}=0$}

When we set $\beta_{-}=0$ we deal with a system where the scale factors $a$
and $b$ in \eqref{Misnvar} are equal. The mini-superspace Lagrangian that
reproduces the field equation assumes the form
\begin{equation} \label{Lag2sc}%
\begin{split}
L= &  \frac{3e^{-3\Omega}}{\kappa N}\left(  \dot{\beta}_{+}^{2}-\dot{\Omega
}^{2}\right)  -\frac{N}{\kappa}\Big[\frac{e^{-8\beta_{+}-\Omega}}{4}%
+\frac{e^{4\beta_{+}-\Omega}}{2}+\frac{\bar{K}}{4}\left(  e^{-2\beta
_{+}-\Omega}+\frac{1}{8}e^{4\beta_{+}-\Omega}\right)  \\
&  +\frac{\bar{\lambda}}{16}\left(  \frac{1}{2}e^{\Omega-4\beta_{+}}%
+e^{2\beta_{+}+\Omega}\right)  +\Lambda e^{-3\Omega}\Big].
\end{split}
\end{equation}

In the case, where $\beta_{-}=0$ we can reduce the system to a single equation
of motion for $\beta_{+}$. To this end, we can solve the constraint equation
algebraically with respect to $N(t)$ and obtain (for $\bar{\lambda}=\Lambda
=0$)
\begin{equation}
\label{lapse2sc}N^{2} = \frac{24 e^{8 \beta_{+} -2 \Omega} \left(  \dot
{\Omega}^{2} - \dot{\beta}_{+}^{2}\right)  }{2 \bar{K} e^{6 \beta_{+} }%
+(\bar{K}+4) e^{12 \beta_{+}}+2}.
\end{equation}
By having satisfied the constraint, the freedom of fixing the gauge passes to
one of the two remaining degrees of freedom, namely $\beta_{+}$ and $\Omega$.
If we choose $\Omega(t)=t$, and as we commented earlier for the more general
case, we are led to an autonomous equation for $\beta_{+}$ that reads
\begin{equation}
\label{redbp}\ddot{\beta}_{+} = \left(  \frac{6 \left(  \bar{K} e^{6 \beta
_{+}}+2\right)  }{2 \bar{K} e^{6 \beta_{+}}+(\bar{K}+4) e^{12 \beta_{+}}+2}+2
\left(  \dot{\beta}_{+} -1\right)  \right)  \left(  1-\dot{\beta}_{+}%
^{2}\right)  .
\end{equation}
The above equation has two obvious solutions $\beta_{+}=0$ and $\beta_{+} =
t$. The first is the one corresponding to the isotropic case and leads to the
corresponding special case expressed by \eqref{soliso1}, the latter is
rejected because it leads to a zero lapse through \eqref{lapse2sc}. We remind
here that the gauge is fixed by setting $\Omega(t)=t$.

Since the equation is autonomous, it has an obvious Lie-point symmetry with
generator $\partial_{t}$ and, as a result, it can be transformed into a first
order non-autonomous Abel equation by taking $\beta_{+}(t) =s$, $t=\int\!
z(s)ds$. The resulting relation is
\begin{equation}
\frac{d z}{ds} = \left(  2 (z-1)-\frac{6 \left(  \bar{K} e^{6 s}+2\right)
z}{2 \bar{K} e^{6 s}+(\bar{K}+4) e^{12 s}+2} \right)  \left(  z^{2}-1 \right)
.
\end{equation}

\paragraph{Evolution of the anisotropic parameter $\beta_{+}\left(  t\right)
$}

Let us now study the evolution of the anisotropic parameter $\beta_{+}\left(
t\right)  $ which satisfies equation (\ref{redbp}). In order to do that, we
study (\ref{redbp}) by numerical integration and its phase space.

\begin{figure}[ptb]
\includegraphics[height=5.4cm]{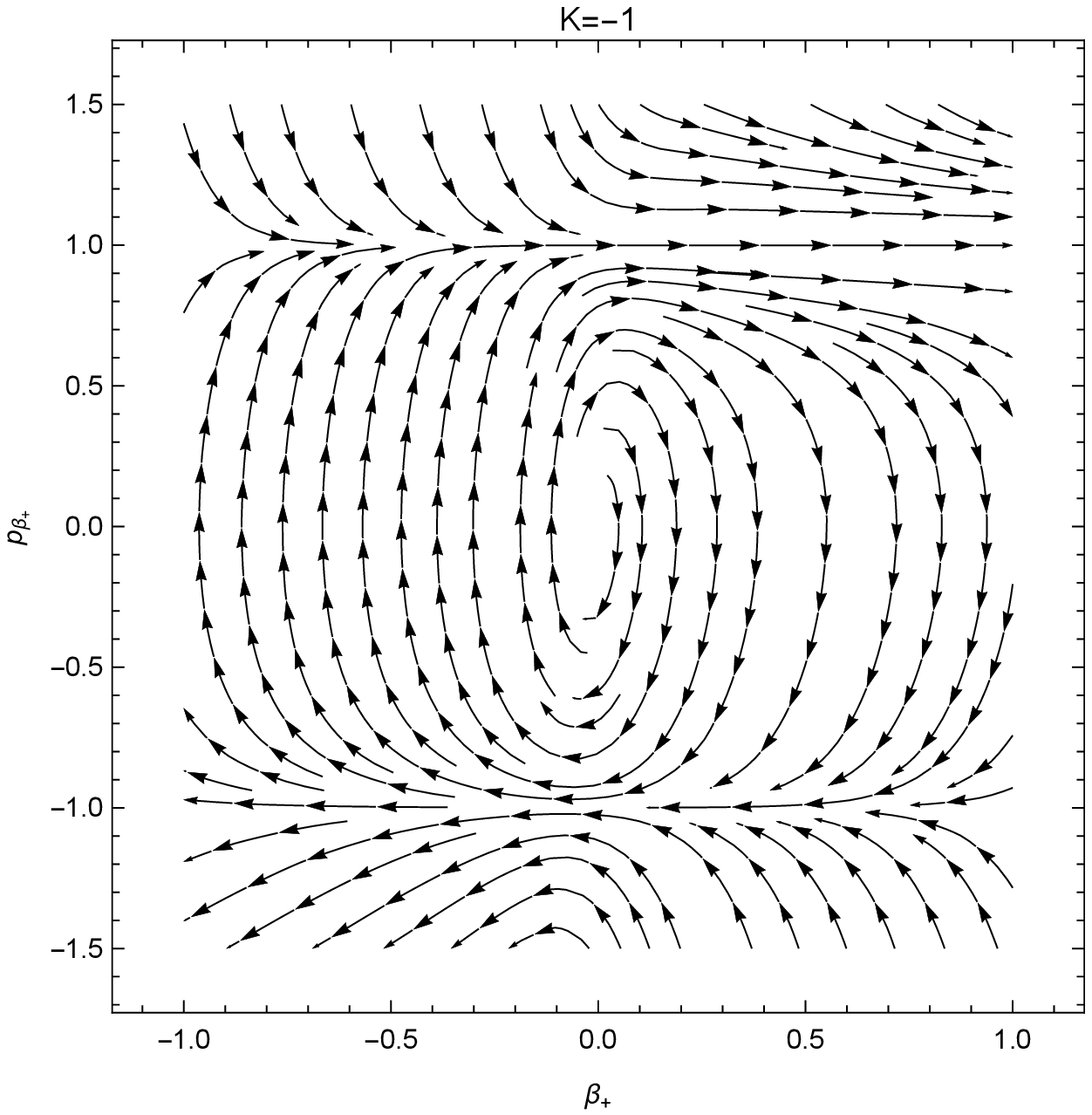}
\includegraphics[height=5.4cm]{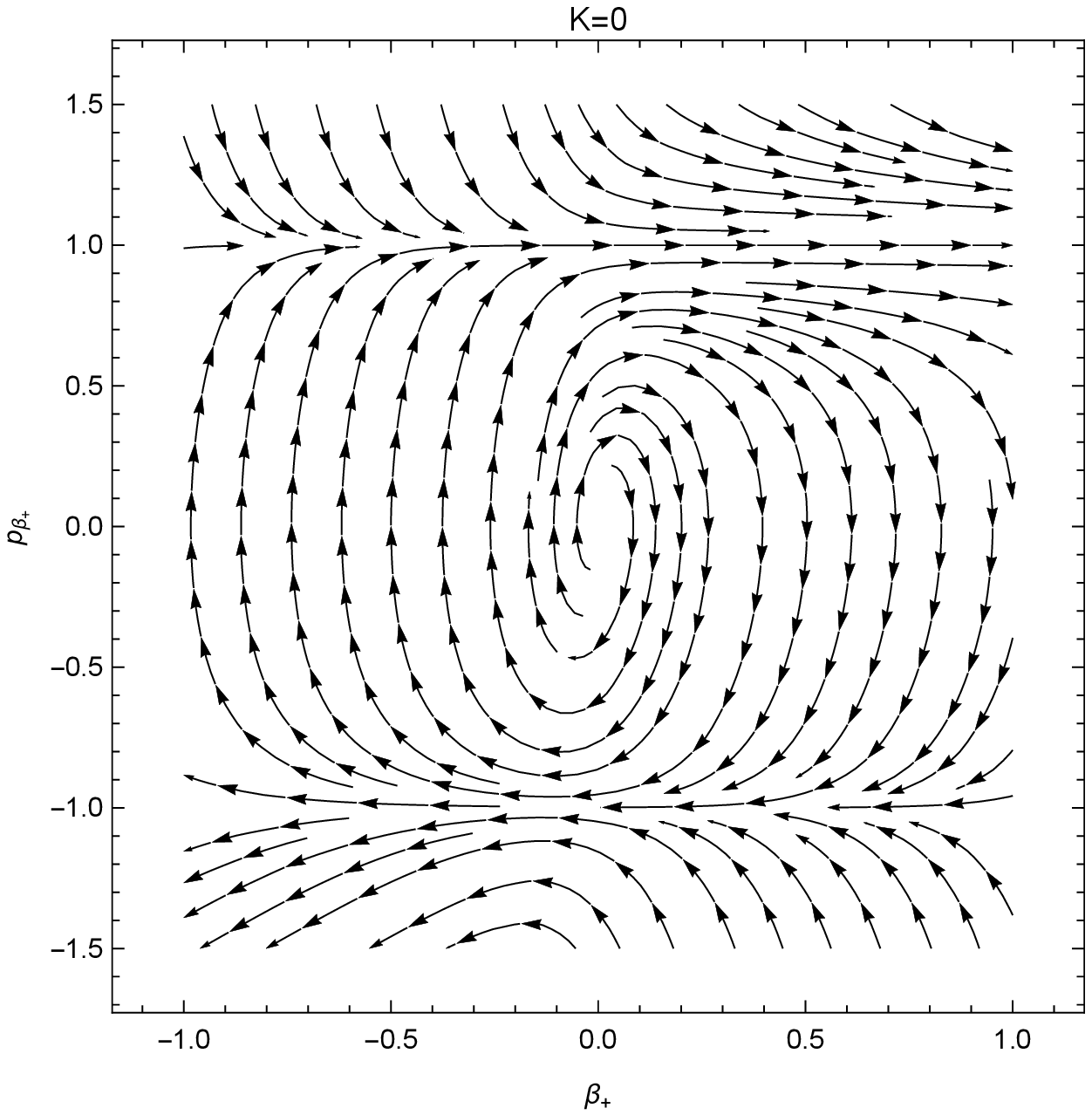}
\includegraphics[height=5.4cm]{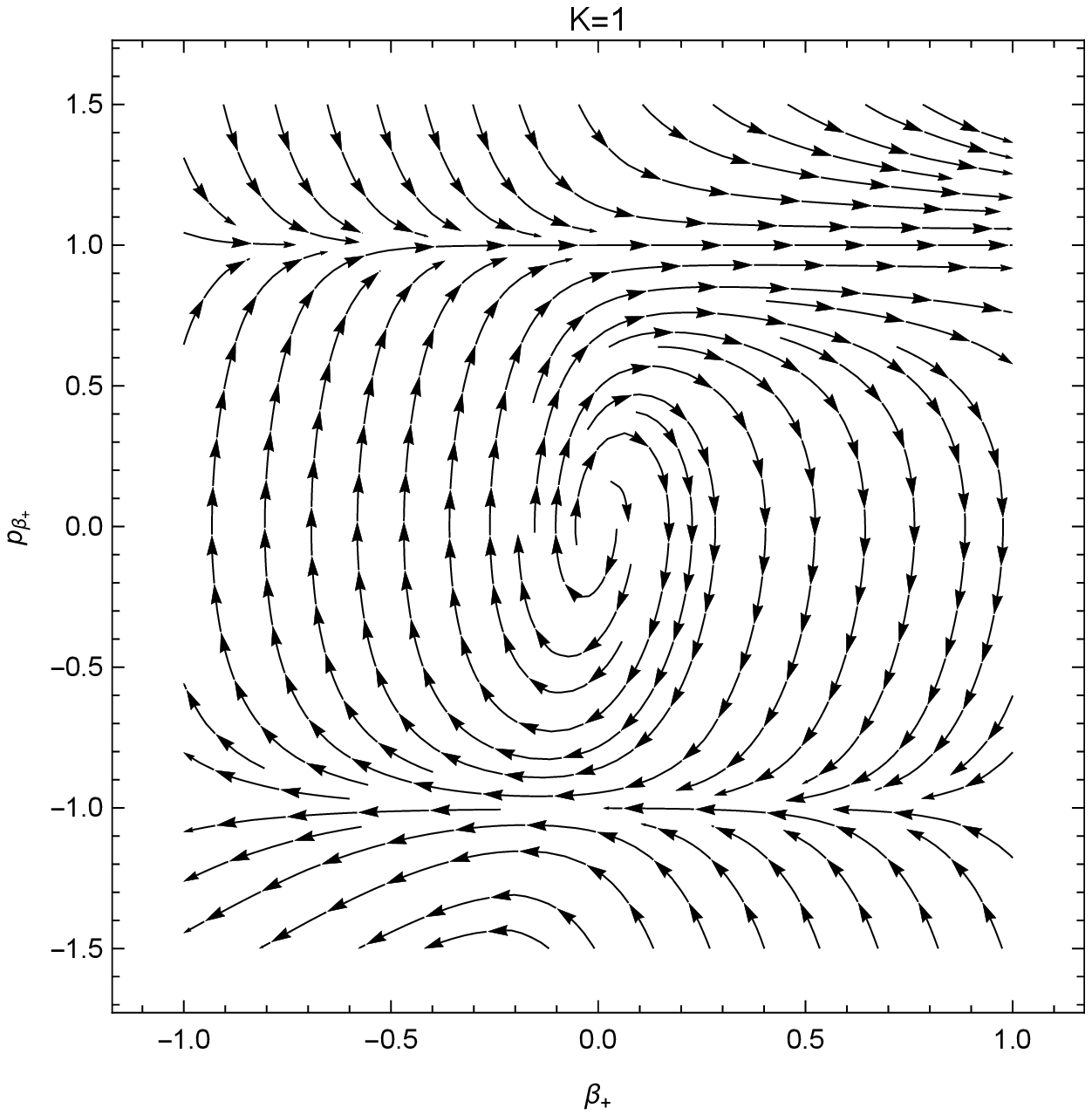}
\caption{Phase portrait of equation (\ref{redbp}) for $\bar{K}$ zero, positive
and negative. We observe that the isotropic universe, i.e. $\beta
_{+}=0,~p_{\beta_{+}}=0$, it is a source and there exists two surfaces as
attractors. Moreover, we observe that the phase space is almost independent
from the value of the parameter $\bar{K}$. }%
\label{phase1}%
\end{figure}

The phase space $\left\{  \beta_{+},p_{\beta+}\right\}  $ of equation
(\ref{redbp}) is presented in in Fig. \ref{phase1} for three different values
of parameter $\bar{K}$, that is, $\bar{K}=-1,~0,~1$, where $p_{\beta+}%
=\dot{\beta}_{+}$. From the diagrams it is clear that the isotropic universe
is an unstable solution, and actually the critical point $\left(  \beta
_{+},p_{\beta+}\right)  =\left(  0,0\right)  $ it is a source for equation
(\ref{redbp}).

\begin{figure}[ptb]
\includegraphics[height=6.5cm]{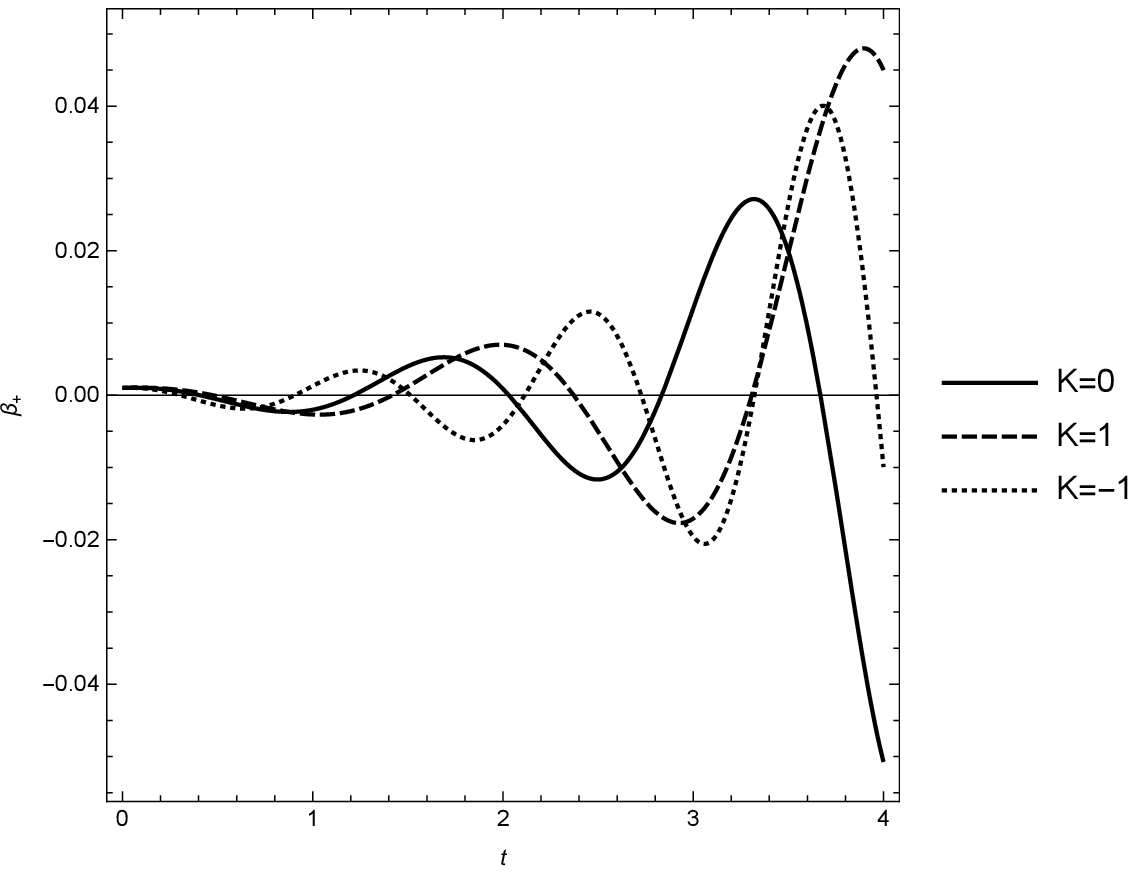}
\includegraphics[height=6.5cm]{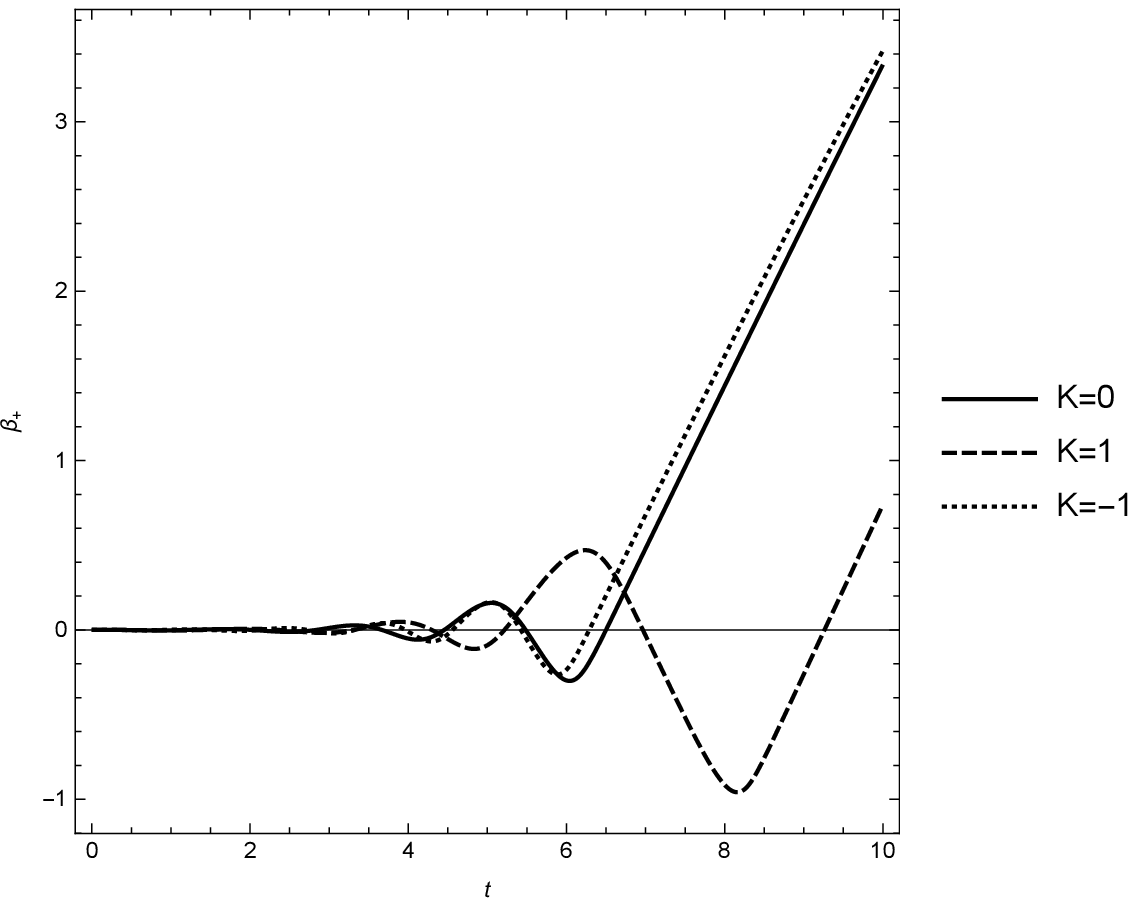}
\caption{Qualitative evolution of the anisotropic parameter $\beta_{+}\left(
t\right)  $ given by the solution of equation (\ref{redbp}) for initial
conditions $\left(  \beta_{+}\left(  0\right)  ,p_{\beta+}\left(  0\right)
\right)  =\left(  10^{-3},10^{-3}\right)  \,$. Solid line is $\bar{K}=0$,
dashed line for $\bar{K}=1$ and dotted line for $\bar{K}=-1$}%
\label{plot1}%
\end{figure}

\begin{figure}[ptb]
\includegraphics[height=6.5cm]{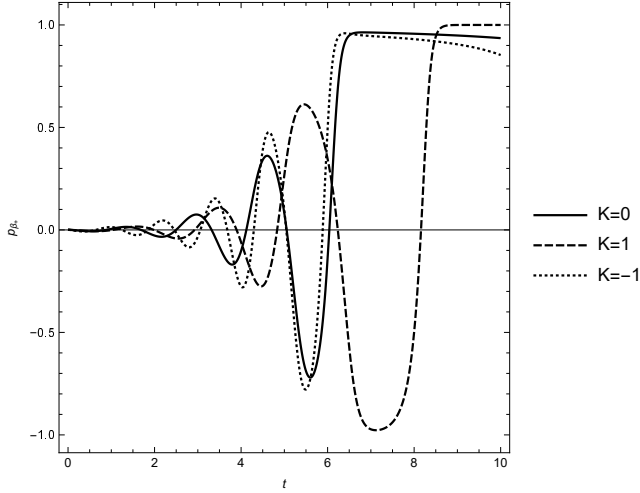}
\caption{Qualitative evolution of the anisotropic parameter $p_{+}\left(
t\right)  $ given by the solution of equation (\ref{redbp}) for initial
conditions $\left(  \beta_{+}\left(  0\right)  ,p_{\beta+}\left(  0\right)
\right)  =\left(  10^{-3},10^{-3}\right)  \,$. Solid line is $\bar{K}=0$,
dashed line for $\bar{K}=1$ and dotted line for $\bar{K}=-1.~$We observer that
$p_{\beta_{+}}\left(  t\right)  $ becomes constant different from zero. }%
\label{plot2}%
\end{figure}

We perform numerical simulations of equation (\ref{redbp}) for initial
conditions close to the critical point $\left(  \beta_{+},p_{\beta+}\right)
=\left(  0,0\right)  $. More specifically we select the initial conditions
$\left(  \beta_{+}\left(  0\right)  ,p_{\beta+}\left(  0\right)  \right)
=\left(  10^{-3},10^{-3}\right)  $ and we perform the numerical simulation for
$\bar{K}=-1,~0,~1$. \ In Fig. \ref{plot1} the evolution of $\beta_{+}\left(
t\right)  $ is presented, where we observe that for the final state of the
solution is an anisotropic system. Furthermore, from Fig. \ref{plot2} where
the evolution of $p_{\beta_{+}}\left(  t\right)  $ is presented, we conclude
that the rate of anisotropy is steady. \

\subsubsection{The special case $\bar{K}=0$, $\bar{\lambda}=0$, $\Lambda=0$}

In respect to the Skyrme coupling, this specific situation that simplifies
significantly \eqref{Lag2sc}, corresponds to $K=\frac{4}{\kappa}$ and
$\lambda=0$. The resulting mini-superspace Lagrangian is
\begin{equation}
L=\frac{3e^{-3\Omega}}{\kappa N}\left(  \dot{\beta}_{+}^{2}-\dot{\Omega}%
^{2}\right)  -\frac{N}{4\kappa}\left(  e^{-8\beta_{+}-\Omega}+2e^{4\beta
_{+}-\Omega}\right)  \label{Lag2scsp}%
\end{equation}
with $N$, $\beta_{+}$ and $\Omega$ being the degrees of freedom of this
problem. Since this results into two spatial equations and one constraint it
means that the physical space (consisting of variables that are independent
and not subjected to a gauge choice) has one degree of freedom. Hence, we
expect that the system can be reduced to a single second order ordinary
differential equation.

It has been shown \cite{Dim2,Dim3}, for systems of this form, there exist
non-local conserved charges that can be constructed out of conformal Killing
vector fields of the mini-superspace metric.

For instance, if we have a singular Lagrangian of the form
\begin{equation}
L=\frac{1}{2N}G_{\alpha\beta}\dot{q}^{\alpha}\dot{q}^{\beta}%
-NV(q),\label{Lagminigen}%
\end{equation}
then a non-local conserved charge
\begin{equation}
I=\xi^{\alpha}p_{\alpha}+\int\!\!N\left(  \omega(q(t))+f(q(t))\right)
V(q(t))dt\label{nonlocgen}%
\end{equation}
exists if $\xi^{\alpha}$ is a conformal Killing vector of the mini-superspace
metric $G_{\alpha\beta}$ with conformal factor $\omega(q)$, i.e.
$\mathcal{L}_{\xi}G_{\alpha\beta}=\omega(q)G_{\alpha\beta}$. The function
$f(q)$ in \eqref{nonlocgen} is the conformal factor of the same vector field
over the potential, $f(q)=\frac{\xi^{\alpha}\partial_{\alpha}V(q)}{V(q)}$ and
$p_{\alpha}$ the momenta $p_{\alpha}=\frac{\partial L}{\partial\dot{q}}$.

In what follows, for convenience and the simplification of the resulting
equations, we make a change in the coordinates of the configuration space, so
as to have the mini-superspace metric in light cone form. Thus, we choose to
reparametrize $\Omega(t)$, $\beta_{+}(t)$ with respect to new variables $u(t)$
and $v(t)$ as
\begin{equation}
\Omega=\frac{u(t)}{3}-v(t),\quad\beta_{+}=\frac{u(t)}{3}+v(t).\label{transc2}%
\end{equation}
The corresponding mini-superspace metric and potential in \eqref{Lagminigen}
are
\begin{equation}
G_{\mu\nu}=\frac{4e^{3v-u}}{\kappa}%
\begin{pmatrix}
0 & 1\\
1 & 0
\end{pmatrix}
\label{mini2sc}%
\end{equation}
and
\begin{equation}
V(u,v)=\frac{e^{-3u-7v}+2e^{u+5v}}{4\kappa}.\label{vec2sc}%
\end{equation}
The equations of motion in the $u,v$ variables read
\begin{subequations}
\label{eq2sc}%
\begin{align}
&  -\frac{4e^{3v-u}\dot{u}\dot{v}}{\kappa N^{2}}-\frac{e^{u+5v}}{2\kappa
}-\frac{e^{-3u-7v}}{4\kappa}=0\label{con2sc}\\
&  \frac{4e^{4v}\dot{N}\dot{v}}{N}-\frac{1}{4}N^{2}e^{-2(u+3v)}\left(
2e^{4(u+3v)}-3\right)  -4e^{4v}\left(  \ddot{v}+3\dot{v}^{2}\right)
=0\label{eqv2sc}\\
&  16e^{2(u+5v)}\dot{N}\dot{u}+N^{3}\left(  7-10e^{4(u+3v)}\right)
+16Ne^{2(u+5v)}\left(  \dot{u}^{2}-\ddot{u}\right)  =0\label{equ2sc}%
\end{align}

In these variables, the configuration space vector $\xi=(-\frac{3}{8},\frac
{1}{8})$ leads to the conformal factors $\omega=\frac{3}{4}$ and $f=\frac
{1}{4}$ over \eqref{mini2sc} and \eqref{vec2sc} respectively. Thus, we can
write the non-local conserved charge as
\end{subequations}
\begin{equation}
I=\frac{e^{3v-u}}{2\kappa N}\left(  \dot{u}-3\dot{v}\right)  +\frac{1}%
{4\kappa}\int\!\!N(t)e^{-3u(t)-7v(t)}+2e^{u(t)+5v(t)}dt.\label{nonloc2sc}%
\end{equation}
It can be easily verified that $\frac{dI}{dt}=0$ whenever the equations of
motion (and the constrained equation) of \eqref{Lagminigen} with
\eqref{mini2sc} and \eqref{vec2sc} are satisfied. Of course the same holds for
the equations of motion of \eqref{Lag2scsp} when we make the inverse
transformation \eqref{transc2}. In principle, it may be assumed that the
constant of motion \eqref{nonloc2sc} may not be of use due to the existence of
the indefinite integral, for the calculation of which one would need to know
the solution in terms of $t$. However, we need to remind to ourselves that at
this point we have not fixed the gauge. If we choose
\begin{equation}
N=\left(  \frac{e^{-3u-7v}+2e^{u+5v}}{4\kappa}\right)  ^{-1},\label{gfix2c}%
\end{equation}
then the originally non-local conserved charge assumes a local form and
becomes
\begin{equation}
I=\frac{e^{3v-u}}{2\kappa N}\left(  \dot{u}-3\dot{v}\right)
+t.\label{nonloc2sc2}%
\end{equation}
In the gauge \eqref{gfix2c}, whatever the solution is, the result of the
integral in the expression \eqref{nonloc2sc} is equal to the variable $t$. As
a result, we have a first order relation that we may use together with the
equations of motion. If we solve algebraically the equation $I=0$%
\footnote{Without loss of generality we can set $I=const.=0$. A non-zero value
for the constant can be absorbed with a time translation since $t$ appears
explicitly only in \eqref{nonloc2sc}. The equations of motion are autonomous
in the gauge \eqref{gfix2c}.} with respect to $\dot{u}$ and substitute into
the constraint equation \eqref{con2sc} (always in the gauge \eqref{gfix2c}),
we can solve the latter with respect to $u(t)$, which yields
\begin{equation}
u(t)=-\frac{1}{4}\ln\left[  \frac{8\kappa^{2}t}{3\dot{v}}-\frac{\kappa^{2}%
}{3\dot{v}^{2}}-2e^{8v}\right]  -v.\label{uofv}%
\end{equation}

Substitution of \eqref{gfix2c} and \eqref{uofv} into the spatial equation
\eqref{eqv2sc} (of course, after the satisfaction of the constraint equation
\eqref{equ2sc} is satisfied when \eqref{eqv2sc} is also satisfied) leads to
the single second order non-autonomous equation
\begin{equation}
\ddot{v}=-\frac{4\dot{v}^{2}\left(  \kappa^{2}-8\kappa^{2}t\dot{v}%
+18e^{8v}\dot{v}^{2}\right)  }{\kappa^{2}\left(  8t\dot{v}-1\right)
}.\label{finv}%
\end{equation}
The latter can be simplified if we set $v=z^{-1/4}$ to
\begin{equation}
8\kappa^{2}\left(  2tz^{3}+\frac{z^{4}}{\dot{z}}\right)  \ddot{z}+\frac
{9\dot{z}^{3}}{z}=0.\label{finv01}%
\end{equation}
It can be seen now, with substitution of \eqref{uofv} and \eqref{gfix2c} into
the constraint \eqref{con2sc} we obtain again \eqref{finv}. The first order
non-local integral, allowed us to extract from the constraint equation the
information about how $u$ is related to $v$ through \eqref{uofv} in the gauge
\eqref{gfix2c}. The special solution $z_{sp}(t)=\frac{3}{4\kappa t}$ leads
through the inverse transformations to $\beta_{+}=0$ and the pure isotropic
case studied earlier.

In terms of $v(t)$ the Ricci scalar becomes
\begin{equation}
R=2e^{-4v}-\frac{\sqrt{3}\dot{v}^{2}\sqrt{-\frac{\kappa^{2}-8\kappa^{2}%
t\dot{v}+6e^{8v}\dot{v}^{2}}{\dot{v}^{2}}}}{\kappa^{2}-8\kappa^{2}t\dot
{v}+6e^{8v}\dot{v}^{2}}. \label{Ricci2sc}%
\end{equation}

It is straightforward to calculate that equation (\ref{finv01}) is invariant
under the action of one parameter point transformation with generator the Lie
symmetry vector
\begin{equation}
X_{L}=t\partial_{t}-z\partial_{t}.
\end{equation}
The latter vector field is responsible for the invariant solution
$z_{sp}(t)=\pm\frac{3}{4\kappa t}.$ \ However it can be used to reduce by one
the order of the differential equation, or write equation (\ref{finv01}) as
time-independent form such that to study the stability of the special solution
$z_{sp}(t)$.

By applying the transformation $z\left(  t\right)  =\frac{Z\left(  t\right)
}{t}$, $t=e^{s}$ in (\ref{finv01}) we remain with the time-independent
second-order differential equation%
\begin{equation}%
\begin{split}
&  -8\left(  k^{2}Z^{4}\left(  2\frac{dZ}{ds}-Z\right)  \right)  \frac{d^{2}%
Z}{ds^{2}}=Z^{4}\left(  9-16k^{2}Z^{2}\right)  \\
&  +\frac{dZ}{ds}\left(  4Z^{3}\left(  14k^{2}Z^{2}-9\right)  +3\frac{dZ}%
{ds}\left(  3\frac{dZ}{ds}\left(  \frac{dZ}{ds}-Z\right)  -2Z^{2}\left(
8k^{2}Z^{2}-9\right)  \right)  \right) \label{ss01}
\end{split}
\end{equation}
which admits the special solution $Z_{sp}\left(  s\right)  =\pm\frac{3}{4k}$.
Hence, by performing the stability analysis of equation (\ref{ss01}) we find
that the special solutions%
\begin{equation}
Z_{sp}\left(  s\right)  =\pm\frac{3}{4k},~Z_{sp}^{\prime}\left(  s\right)  =0
\end{equation}
are "attractors" for the solution of equation (\ref{ss01}).

Consider the symmetry vector $X_{L}$, from there we can define the invariant
functions $v=zt~\ ,~u=t^{2}\dot{z}$, which we use to reduce the differential
equation in the following rational first-order equation%
\begin{equation}
-8k^{2}v^{4}\left(  v+u\right)  \left(  v+2u\right)  \frac{du}{dv}=u\left(
9u^{3}-32k^{2}v^{4}u-16k^{2}v^{5}\right)  ,
\end{equation}
whose solution can not be written in a closed form expression.

Let us now use the method of singularity analysis to write the solution of
equation (\ref{finv01}) in an algebraic form. That is feasible because of the
existence of the singular solution $z_{sp}\left(  t\right)  $. We proceed, by
increasing by one order of equation (\ref{finv01}) with the change of
variables%
\begin{equation}
t=\frac{1}{Y\left(  \tau\right)  }~,~z=-\frac{1}{Y^{2}}\frac{dY}{d\tau}.
\label{ss02}%
\end{equation}

The resulting equation is a third-order, time-dependent differential equation
which admits the singular solution $Y\left(  \tau\right)  =Y_{0}\left(
\tau-\tau_{0}\right)  ^{-1}$. That is a singular solution which means that a
movable singularity exists for the third-order differential equation. We
continue by applying the ARS algorithm \cite{ars1} and we find that the
resonances are $r=-1,0,1$.

Hence, we find that equation (\ref{finv01}) under the change of variables
(\ref{ss02}) passes the singularity test, and the algebraic solution is given
by the Right Painlev\'{e} Series
\begin{equation}
Y\left(  \tau\right)  =Y_{0}\left(  \tau-\tau_{0}\right)  ^{-1}%
+\displaystyle\sum\limits_{i=1}^{\infty}Y_{i}\left(  \tau-\tau_{0}\right)
^{-1+i}%
\end{equation}
with constants of integrations the parameters $Y_{0,}~\tau_{0}$ and $Y_{1}$,
while first two nonzero coefficients $Y_{2}$ and $Y_{3}$ are derived to be%
\begin{equation}
Y_{2}=\frac{Y_{1}^{2}}{3k^{2}Y_{0}}\left(  9Y_{0}^{2}Y_{1}^{2}+k^{2}\right)
~,~Y_{3}=\frac{3Y_{1}^{5}}{2k^{4}}\left(  7k^{2}+18\left(  Y_{0}Y_{1}\right)
^{2}\right)  .
\end{equation}

We conclude that the field equations in the LRS case are integrable; for the
application of the ARS algorithm in the Mixmaster Universe we refer the reader
in \cite{Intp0,Intp0a}.

\section{Quantum description}

\label{Sec6}

\subsection{Hamiltonian Formalism and Quantum description}

By having as a starting point Lagrangian \eqref{miniLag} we can proceed to
write the Hamiltonian for the system with the use of the Dirac Bergmann
algorithm \cite{Dirac,Berg}. Without getting into details, the well known
result in this case is that you obtain a Hamiltonian $H$ which is a linear
combination of the constraints $p_{N} \approx0$ and $\mathcal{H} \approx
0$\footnote{The ``$\approx$" symbol denotes a weak equality, i.e. quantities
which are zero themselves, but whose gradients with respect to phase space
variables are not.}
\begin{equation}
H = N \mathcal{H} + u_{N} p_{N} .
\end{equation}
The function $u_{N}$ is arbitrary and is not essential for the theory. In our
case the Hamiltonian constraint $\mathcal{H}$ reads
\begin{equation}
\mathcal{H} = \frac{\kappa e^{3\Omega}}{12} \left(  p_{\beta_{+}}^{2} +
p_{\beta-}^{2} - p_{\Omega}^{2}\right)  + V(\beta_{+},\beta_{-},\Omega),
\end{equation}
where $p_{\beta_{+/-}} = \frac{\partial L}{\partial\dot{\beta}_{+/-}}$ and
$p_{\Omega}= \frac{\partial L}{\partial\dot{\Omega}}$ are the momenta of the system.

By making the typical identification
\begin{equation}
X\mapsto\widehat{X}=X,\quad p_{X}\mapsto\widehat{p}_{X}=-\mathbbmtt{i}{i}\frac
{\partial}{\partial X}, \label{mapop}%
\end{equation}
where $X$ may be $\Omega,\beta_{+},\beta_{-}$ or $N$, we can proceed with the
canonical quantization for the system. Dirac's prescription for constrained
system dictates to demand that the wave function, $\Psi$, is invariant under
the action of the constraints, i.e. $\widehat{p}_{N}\Psi=0$ and $\widehat
{\mathcal{H}}\Psi=0$. The first signifies that $\Psi$ does not depend on the
variable $N$, while the second is the well-known Wheeler-DeWitt equation. In
order to address the factor ordering problem raised by the kinetic term of
$\mathcal{H}$ we choose to use the conformal Laplacian operator
\[
\widehat{L}=\nabla_{\mu}\nabla^{\mu}+\frac{d-2}{4(d-1)}\mathcal{R},
\]
where $\mathcal{R}$ is the Ricci scalar of the mini-superspace metric that we
read from \eqref{miniLag} and $d$ is its dimension.\footnote{The Greek indices
appearing in this section do not dictate space-time coordinates, but
correspond to the configuration space variables $\Omega,\beta_{+}$ and
$\beta_{-}$. The covariant derivatives $\nabla_{\mu}$ are calculated with
respect to the mini-superspace metric.} The mini-superspace metric in our case
is $\mathrm{G}_{\mu\nu}=\frac{6}{\kappa}e^{-3\Omega}\mathrm{diag}(-1,1,1)$. As
a result, we can write
\begin{equation}
\widehat{\mathcal{H}}=-\frac{1}{2}\widehat{L}+V(\beta_{+},\beta_{-}%
,\Omega)=-\frac{1}{2}\nabla_{\mu}\nabla^{\mu}-\frac{1}{16}\mathcal{R}%
+V(\beta_{+},\beta_{-},\Omega), \label{conquant}%
\end{equation}
where $\mathcal{R}=\frac{3\kappa}{4}e^{3\Omega}$.

In this way $\widehat{\mathcal{H}}$ is a linear, Hermitian (under appropriate
boundary conditions) operator. Additionally, due to the fact of using the
conformal Laplacian, the classical symmetry of arbitrary reparametrizing the
lapse function can be seen at the quantum level as a conformal transformation
of the mini-superspace metric. What is more, the probability amplitude
\[
\rho\sim\sqrt{-\mathrm{G}}\Psi^{\ast}\Psi d\Omega d\beta_{+}d\beta_{-},
\]
where $G$ is the determinant of the mini-superspace metric, transforms as a
scalar under coordinate changes in the configuration space variables. With the
help of \eqref{conquant} the Wheeler-DeWitt equation can be written as
\begin{equation}
-\frac{\kappa e^{3\Omega}}{12}\left(  \frac{\partial^{2}\Psi}{\partial
\beta_{+}^{2}}+\frac{\partial^{2}\Psi}{\partial\beta_{-}^{2}}-\frac
{\partial^{2}\Psi}{\partial\Omega^{2}}+\frac{3}{2}\frac{\partial\Psi}%
{\partial\Omega}\right)  -\frac{3\kappa e^{3\Omega}}{64}\Psi+V(\beta_{+}%
,\beta_{-},\Omega)\Psi=0.
\end{equation}

\subsection{Quantum description in the Isotropic case}

In this section we proceed with the quantum description of this
one-dimensional system. The problem can be simplified if we perform a
reparametrization of the lapse function $N\mapsto n=NV(\Omega)$, where
$V(\Omega)$ is the potential part of Lagrangian \eqref{Lag1sc}. In this way
all the information of the system is passed into the kinetic term and the new
equivalent Lagrangian is
\begin{equation}
L=\frac{1}{2n}h(\Omega)\dot{\Omega}^{2}-n,
\end{equation}
where
\begin{equation}
h(\Omega)=-\frac{3e^{-6\Omega}\left(  12(\bar{K}+2)e^{2\Omega}+32\Lambda
+3\bar{\lambda}e^{4\Omega}\right)  }{16\kappa^{2}} \label{h1sc}%
\end{equation}
The canonical quantization of a pure gauge system of this form has been
performed in \cite{Dim1}. By following the Dirac-Bergmann algorithm
\cite{Dirac,Berg} we are led to the classical Hamiltonian
\begin{equation}
H=n\mathcal{H}+u_{n}p_{n},
\end{equation}
where $u_{n}$ is an arbitrary function, while $p_{n}\approx0$ and
$\mathcal{H}=\frac{1}{2h(\Omega)}p_{\Omega}^{2}+1\approx0$ are the constraints
of the system, whose quantum analogs are to annihilate the wave function. As
previously, we introduce the mapping
\[
\Omega\mapsto\widehat{\Omega}=\Omega,\quad n\mapsto\widehat{n}=n,\quad
p_{\Omega}\mapsto\widehat{p}_{\Omega}=-\mathbbmtt{i}\frac{\partial}%
{\partial\Omega},\quad p_{n}\mapsto\widehat{p}_{n}=-\mathbbmtt{i}\frac
{\partial}{\partial n}%
\]
and according to Dirac's prescription quantizing constrained system we require
$\widehat{p}_{n}\Psi=0$ and $\widehat{\mathcal{H}}\Psi=0$. For the kinetic
term of the Hamiltonian we use the one-dimensional equivalent of the
Laplacian, i.e the Hermitian operator
\begin{equation}
\widehat{\mathcal{H}}=-\frac{1}{2\mu(\Omega)}\frac{d}{d\Omega}\left(
\frac{\mu(\Omega)}{h(\Omega)}\frac{d}{d\Omega}\right)  +1, \label{Hamop1sc}%
\end{equation}
where $h(\Omega)$ is given by \eqref{h1sc} and $\mu(\Omega)=(-h(\Omega
))^{1/2}$ is the measure function. Under operator \eqref{Hamop1sc} the
solution to the Wheeler-DeWitt equation $\widehat{\mathcal{H}}\Psi(\Omega)=0$
is \cite{Dim1}
\begin{equation}
\Psi(\Omega)=C_{1}e^{\mathbbmtt{i}{i}\sqrt{2}\int\sqrt{-h(\Omega)}d\Omega
}+C_{2}e^{-\mathbbmtt{i}{i}\sqrt{2}\int\sqrt{-h(\Omega)}d\Omega},
\end{equation}
where $C_{1}$, $C_{2}$ are constants of integration.

\subsection{Two scale factors ($\beta_{-}=0$).}

In the case where we take $\beta_{-}=0$, the Dirac-Bergmann algorithm yields
the Hamiltonian constraint
\begin{equation}
\label{Hcon2d}\mathcal{H} = \frac{1}{12} e^{3 \Omega} \left(  p_{\beta_{+}%
}^{2} - p_{\Omega}^{2}\right)  + V_{2d} \approx0,
\end{equation}
where
\begin{equation}%
\begin{split}
\label{V2d}V_{2d} =  &  \frac{1}{2 \kappa} \left(  \frac{e^{-8 \beta
_{+}-\Omega}}{2} + e^{4 \beta_{+} -\Omega} \right)  + \frac{\Lambda e^{-3
\Omega}}{\kappa} + \frac{\bar{K}}{4 \kappa} \left(  e^{-2 \beta_{+} -\Omega} +
\frac{e^{4 \beta_{+} -\Omega}}{2} \right) \\
&  + \frac{\bar{\lambda}}{16 \kappa} \left(  e^{2\beta_{+} +\Omega} +
\frac{e^{\Omega-4 \beta_{+}}}{2} \right)
\end{split}
\end{equation}
is the corresponding potential for the two dimensional case.

By putting in use the Laplacian so as to express the kinetic term of
\eqref{Hcon2d} in its quantum counterpart $\widehat{H}=-\frac{1}{2}\nabla
_{\mu}\nabla^{\mu}+ V_{2d}$, the Wheeler-DeWitt equation $\widehat{H}\Psi=0$
reads
\begin{equation}
\frac{\kappa e^{3 \Omega}}{12} \left(  \frac{\partial^{2} \Psi}{\partial
\Omega^{2}}- \frac{\partial^{2} \Psi}{\partial\beta_{+}^{2}}\right)  + V_{2d}
\Psi=0
\end{equation}

\subsubsection{The $\bar{K}=\bar{\lambda}=\Lambda=0$ case}

The corresponding Wheeler-DeWitt equation, under the conditions $\bar{K}%
=\bar{\lambda}=\Lambda=0$, is
\begin{equation}
\frac{\kappa e^{3\Omega}}{12}\left(  \frac{\partial^{2}\Psi}{\partial
\Omega^{2}}-\frac{\partial^{2}\Psi}{\partial\beta_{+}^{2}}\right)  +\frac
{1}{2\kappa}\left(  \frac{1}{2}e^{-8\beta_{+}-\Omega}+e^{4\beta_{+}-\Omega
}\right)  \Psi=0.
\end{equation}

\section{Conclusions and perspectives}

\label{Sec7}

The dynamical properties of the Bianchi IX cosmology with three independent
scale factors for the Einstein-$SU(2)$ Skyrme have been studied. The
generalized hedgehog approach in a sector with unit Baryonic charge provides
with an ansatz for the Skyrmion with the remarkable property that the
\textit{matter field equations are automatically satisfied for any Bianchi IX
metric}. This allows to find non-trivial exact solutions in which the Bianchi
IX cosmology is sourced by a topological soliton. Due to the fact that the
complete set of coupled Einstein-Skyrme field equations in the above mentioned
non-trivial topological sector can be reduced to three dynamical equations for
the three Bianchi IX scale factors, one can derive a mini-superspace action.
This important property allows to analyze in details the classical
integrability properties of the Bianchi IX metric in the Einstein-Skyrme
system. Especially it has been shown that the LRS case is integrable.
Moreover, it is possible to derive the Wheeler de-Witt equation for the
Bianchi IX metric in the Einstein-Skyrme cosmology in which all the effects of
the Skyrmion are encoded in an effective potential. These results open the
possibility to study the cosmological consequences of topological solitons in
a model which is very relevant from the phenomenological point of view.

\subsection*{Acknowledgements}

This work has been funded by the Fondecyt grants 1160137, 1150246, 3160121.
The Centro de Estudios Cient\'{\i}ficos (CECs) is funded by the Chilean
Government through the Centers of Excellence Base Financing Program of Conicyt.

\appendix

\section{}

\label{appRmu}

The three four vectors that make up $R_{\mu}=R_{\mu}^{i}t_{i}$ are calculated
to be
\begin{subequations}
\label{Rmus}%
\begin{align}
R_{\mu}^{1} &  =(0,\frac{1}{2}\sin\phi,0,-\frac{1}{2}\sin\theta\cos\phi)\\
R_{\mu}^{2} &  =(0,-\frac{1}{2}\cos\phi,0,-\frac{1}{2}\sin\theta\sin\phi)\\
R_{\mu}^{3} &  =(0,0,\frac{1}{2},\frac{1}{2}\cos\theta).
\end{align}
The $t_{i}$ satisfy an algebra isomorphic to the one constructed by the
$1$-forms $\omega^{\alpha}$ in \eqref{Cartanforms}
\end{subequations}
\begin{equation}
\lbrack t_{i},t_{j}]=-2\epsilon_{ijk}t_{k},
\end{equation}
which explains the compatibility of the equations of motion. It is well known
that there is a (two to one) correspondence between the $SU(2)$ group related
with the matter content and the $SO(3)$ group of the space-time isometries.
What is more, with the ansatz for the Skyrme field that we adopt here, the
three $R_{\mu}^{i}$ of \eqref{Rmus} are multiples of the Cartan forms
$\omega^{\alpha}$ \eqref{Cartanforms}, with an interchange in the $\gamma$,
$\phi$ variables. This difference in $\gamma$ and $\phi$ is not of importance,
we could have taken our ansatz with $\gamma\leftrightarrow\phi$ and be led to
the same results but with a topological charge $B=-1$ instead of $B=+1$.

\section{}

Here we discuss why the Skyrme field equations are identically satisfied (with
the ansatz defined in Eqs. (\ref{bans0}), (\ref{bans1}), (\ref{bans2}),
(\ref{bSkans}) and (\ref{bSkans2}) on any Bianchi IX metric identically.

Let us consider the following slightly modified ansatz the $SU(2)$ scalar
$U(x^{\mu})$:
\begin{equation}
U^{\pm1}=Y^{0}(x^{\mu})\mathbb{I}\pm Y^{i}(x^{\mu})t_{i},\quad(Y^{0}%
)^{2}+(Y^{i})^{2}=1 \label{Aans0}%
\end{equation}
where $\mathbb{I}$ is the two dimensional identity matrix and $Y^{\mu}$ is
parametrized as
\begin{equation}
Y^{0}=\cos\alpha,\quad Y^{i}=n^{i}\sin\alpha\label{Aans1}%
\end{equation}
where
\begin{equation}
n^{1}=\sin\Theta\cos\Phi,\quad n^{2}\sin\Theta\sin\Phi,\quad\cos\Theta.
\label{Aans2}%
\end{equation}
In what follows we denote the space-time coordinates with the variables
$x^{\mu}=(t,\theta,\phi,\gamma)$ and adopt the following ansatz for the
Skyrmion (see \cite{canfora6} \cite{canfora6b} \cite{canfora6c} and references
therein)
\begin{equation}
c_{1}\Phi=\frac{\gamma+\phi}{2},\quad\tan\left(  c_{2}\Theta\right)
=\frac{\cot\left(  \frac{\theta}{2}\right)  }{\cos\left(  \frac{\gamma-\phi
}{2}\right)  },\quad\tan\left(  c_{3}\alpha\right)  =\frac{\sqrt{1+\tan
^{2}\Theta}}{\tan\left(  \frac{\gamma-\phi}{2}\right)  }\ , \label{aSkans}%
\end{equation}%
\begin{equation}
R_{\mu}=U^{-1}\partial_{\mu}U=R_{\mu}^{i}t_{i}\ ,\ \ c_{j}\in%
\mathbb{R}
\ , \label{aSkans2}%
\end{equation}
where the $c_{j}$ are real parameters to be determined solving the Skyrme
field equations on a generic Bianchi IX space-time with three independent
scale factors.

Then, one can see that the three Skyrme field equations in Eq. (\ref{Skeq})
with the ansatz defined in Eqs. (\ref{Aans0}), (\ref{Aans1}), (\ref{Aans2}),
(\ref{aSkans}) and (\ref{aSkans2}) are satisfied \textit{if and only if}%
\[
c_{j}=1\ ,\ j=1,2,3\ ,
\]
namely%
\[
\Sigma^{j}=0\ \ \forall j=1,2,3\ \Leftrightarrow\ c_{j}=1\ ,\ j=1,2,3.
\]
Thus, the only possible ansatz with the properties that the Skyrme field
equations are satisfied identically in the most general Bianchi IX metric is
the one in Eqs. (\ref{Aans0}), (\ref{Aans1}), (\ref{Aans2}), (\ref{aSkans})
and (\ref{aSkans2}) with $c_{j}=1$ for $j$=$1$,$\ 2$,$\ 3$ (which, of course,
is the one considered in the main text).

\end{document}